\documentclass{mem}
\usepackage{natbib}
\usepackage{txfonts}
\usepackage{balance}
\usepackage{graphicx}
\usepackage[a4paper]{hyperref}
\idline{75}{282}

\newcommand{\chem}[2]{$\mathrm{^{#1}#2}$}
\defcitealias{costa2006}{C06}
\defcitealias{pumo2006}{P06}
\defcitealias{pumo2010}{P10}

\begin{document}
\title{Uncertaities due to convection and production of s-nuclei in massive stars}

\author{
M.L. \,Pumo\inst{1,2} 
\and G. \,Contino\inst{3}
\and A. \,Bonanno\inst{2}
\and R.A. \,Zappal\`a\inst{3}
}

\offprints{M.L. Pumo}

\institute{
INAF - Osservatorio Astronomico di Padova, Vicolo dell'Osservatorio 5, I-35122
Padova, Italy
\and
INAF - Osservatorio Astrofisico di Catania, Via S. Sofia 78,
I-95123 Catania, Italy
\and
Universit\`a di Catania,
Dipartimento di Fisica e Astronomia (Sez. astrofisica), Via S. Sofia 78,
I-95123 Catania, Italy
}

\authorrunning{Pumo et al.}

\titlerunning{Uncertaities due to convection and production of s-nuclei in massive stars}
   \subtitle{}

\abstract{
We have investigated the role of the convective overshooting on the production of s-nuclei during core He-burning phase in stellar models with different initial mass and metallicity ($15 \leq M_{ZAMS}/M_{\odot} \leq 25$; $10^{−4} \leq Z \leq 0.02$), considering a diffusive approach to model the convective overshooting. The results and their possible consequences on some open questions linked to the s-process weak component efficiency are discussed.

}

\maketitle


\section{Introduction}

As first pointed out by \citet{b2fh}, about half of the elements between Fe and Bi are formed via the so-called s-process, through neutron capture reactions and beta decays along the ``valley of stability''. 

In order to explain the observed solar distribution of these elements (so-called s-nuclei), more than one s-process ``component'' (i.e. event with a single set of physical conditions like neutron exposure, initial abundances and neutron density) is necessary, and current views on the subject suggest the existence of two components: the so-called main and weak component of s-process, respectively. These components, in terms of stellar environments, correspond to two distinct categories of stars in different evolutionary phases \citep[e.g.][]{kappeler99}. In particular, the main component occurs in low mass stars ($M_{ZAMS} \sim 1.5$-$3 M_\odot$) during their asymptotic giant branch (AGB) phase, and the main neutron source is the $^{13}$C($\alpha$,n)$^{16}$O reaction; while the other one takes place in massive stars ($M_{ZAMS} \gtrsim 13 M_{\odot}$) primarily during their core He-burning phase, and the main neutron source is the $^{22}$Ne($\alpha$,n)$^{25}$Mg reaction.

In addition to these two components, other kinds of stars, as massive AGB ($M_{ZAMS} \sim 4$-$7 M_\odot$) and super-AGB stars ending their evolution as NeO white dwarfs ($M_{ZAMS} \sim 7.5$-$10 M_\odot$; for details see e.g. Fig. 1 in \citealt{pumo2009b}, but also \citealt{ps07} or \citealt{pumo07} and references therein), could also contribute to synthesize s-nuclei, but this hypothesis still needs further investigation \citep[][and references therein]{pumo2009a}. Moreover, in some studies \citep[see e.g.][]{gallino98,lugaro03} is also suggested the existence of a ``strong'' component, taking place in low-metallicity stars of low-intermediate mass during AGB phase, which should be responsible for the synthesis of s-nuclei around the \chem{208}{Pb}. Furthermore, \citet[][]{travaglio04} propose the existence of an additional component referred as lighter element primary s-process (LEPP), but its nature is still unclear and under debate \citep*[e.g.][and references therein]{tur09,pignatari10}.

Concerning the weak component, the general features of this nucleosynthesis process seem to be well established, but there are still some open questions linked to the nuclear physics, the stellar evolution modelling, and the possible contribution to the s-nucleosynthesis from post-He-burning stellar evolutionary phases (see e.g. \citealt{woosley2002}; \citealt{pumo2006}, hereafter P06; \citealt{costa2006}, hereafter C06).\par

Many works \citep[see e.g.][]{arcoragi1991,rayethashimoto2000,the2000,the07,hoffman2001,raucher02,LM03,tur07,tur09,pignatari10,bennett10} have been devoted to analyze these uncertainties. However the treatment of convective overshooting, which represents one of the major uncertainties due to the stellar evolution modelling, makes exceptions because its role on the production of s-nuclei in massive stars has not been intensively investigated (see e.g. \citetalias{pumo2006} and references therein), despite convective overshooting can affect the efficiency of the s-process by giving rise to an increase of the amount of material that experiences neutron irradiation and to a variation of the s-process lifetime (for details see e.g. \citetalias{costa2006}). 

In the light of our preparatory studies on this topic (\citetalias{pumo2006} and \citetalias{costa2006}), which show a not negligible impact of the convective overshooting on the s-process during core He-burning in stellar models with an initial (i.e. at ZAMS) mass of $25 M_\odot$ and an initial metallicity of $Z=0.02$, we believe it is worthwhile examining this issue further by analyzing the s-process efficiency in stellar models with different mass and metallicity, using several values for the overshooting parameter $f$, which determines the overall efficiency of convective overshooting when a diffusive approach is used to model it (see \citetalias{costa2006} and references therein for details on the diffusive approach).


\section{Input physics and models}

We performed 22 s-process simulations considering two grids of stellar models having different initial masses ($15 \leq M_{ZAMS}/M_{\sun} \leq 25$), initial metallicities ($10^{−4} \leq Z \leq 0.02$) and overshooting parameter values ($f= 10^{-5}$ for models without overshooting, and $f= 0.01, 0.02$ and $0.035$ for models with overshooting).

These simulations have been performed through the ``post-processing'' technique \citep[see e.g.][]{prantzos87}, using the stellar evolution code, the s-nucleosynthesis code and the s-process network described in \citetalias{pumo2006} and \citetalias{costa2006}, with the differences reported in \citet[][]{pumo2010}, hereafter P10.


\section{Results and discussion}

\begin{table*}
\caption{\footnotesize Parameters describing the s-process efficiency (see text) for the $Z=0.02$ stellar models with $M_{ZAMS}= 15 M_\odot$ (a), $M_{ZAMS}= 20 M_\odot$ (b) and $M_{ZAMS}= 25 M_\odot$ (c). The overshooting parameter value is reported in the first column for each set of stellar models having a fixed initial mass. (Table adapted from \citetalias{pumo2010}).}
\label{tab_Zcost}
\begin{center}
\begin{tabular}{l l c c c c}
  \hline\hline
      &$f$       &$F_0$     &$n_c$  &$MCZME$   &$Duration$ [$sec$] \\
  \hline
  (a) &$10^{-5}$ &$9.80 $   &$1.19$ &$1.89M_{\odot}$&$5.44\times10^{13}$ \\
      &$0.01$    &$15.45$   &$1.80$ &$2.54M_{\odot}$&$5.42\times10^{13}$ \\
      &$0.02$    &$27.32$   &$2.50$ &$2.90M_{\odot}$&$5.16\times10^{13}$ \\
      &$0.035$   &$55.96$   &$3.35$ &$3.56M_{\odot}$&$4.40\times10^{13}$ \\
 \hline
  (b) &$10^{-5}$ &$43.85 $  &$3.03$ &$3.26M_{\odot}$&$3.77\times10^{13}$ \\
      &$0.01$    &$49.31 $  &$3.19$ &$3.94M_{\odot}$&$3.73\times10^{13}$ \\
      &$0.02$    &$90.10 $  &$3.90$ &$4.41M_{\odot}$&$3.65\times10^{13}$ \\
      &$0.035$   &$172.56$  &$4.74$ &$4.81M_{\odot}$&$3.59\times10^{13}$ \\
 \hline
  (c) &$10^{-5}$ &$92.92 $  &$3.96$ &$5.40M_{\odot}$&$2.32\times10^{13}$ \\
      &$0.01$    &$164.72$  &$4.68$ &$6.48M_{\odot}$&$2.13\times10^{13}$ \\
\end{tabular}
\end{center}
\end{table*}
\begin{table*}
\caption{\footnotesize As in Table \ref{tab_Zcost}, but for the $M_{ZAMS}= 20 M_\odot$ stellar models with $Z=10^{-4}$ (a), $Z=0.005$ (b) and $Z=0.01$ (c). (Table adapted from \citetalias{pumo2010}).}
\label{tab_Mcost}
\begin{center}
\begin{tabular}{l l c c c c}
  \hline\hline
      &$f$        &F$_0$     &n$_c$   &MCZME            &Duration [$sec$]  \\
  \hline
  (a) &$10^{-5}$  &$3.78 $   &$0.16$  &$3.52M_{\odot}$  &$3.58\times10^{13}$  \\
      &$0.01$     &$4.30 $   &$0.23$  &$4.54M_{\odot}$  &$3.57\times10^{13}$  \\
      &$0.02$     &$4.27 $   &$0.23$  &$5.19M_{\odot}$  &$3.18\times10^{13}$  \\
      &$0.035$    &$4.34 $   &$0.24$  &$5.58M_{\odot}$  &$3.08\times10^{13}$  \\
 \hline
  (b) &$10^{-5}$  &$9.56  $  &$1.19$  &$3.96M_{\odot}$  &$4.04\times10^{13}$  \\
      &$0.01$     &$10.47 $  &$1.31$  &$4.46M_{\odot}$  &$3.78\times10^{13}$  \\
      &$0.02$     &$11.06 $  &$1.40$  &$4.93M_{\odot}$  &$3.21\times10^{13}$  \\
      &$0.035$    &$11.14 $  &$1.40$  &$5.26M_{\odot}$  &$2.89\times10^{13}$  \\
 \hline
  (c) &$10^{-5}$  &$48.05 $  &$3.08$  &$3.94M_{\odot}$  &$4.09\times10^{13}$  \\
      &$0.01$     &$72.14 $  &$3.65$  &$4.15M_{\odot}$  &$3.79\times10^{13}$  \\
      &$0.02$     &$68.27 $  &$3.57$  &$4.80M_{\odot}$  &$3.40\times10^{13}$  \\
      &$0.035$    &$125.44$  &$4.29$  &$5.23M_{\odot}$  &$3.14\times10^{13}$  \\
 \hline
\end{tabular}
\end{center}
\end{table*}

The results are summarized in Tables \ref{tab_Zcost} and \ref{tab_Mcost} in terms of the following parameters, that describe the s-process efficiency: 
\begin{itemize}
 \item[(a)] the average overproduction factor F$_0$ for the 6 s-only nuclei within the mass range $70 \leq A \leq 87$;
 \item[(b)] the number of neutrons captured per \chem{56}{Fe} seed nucleus n$_c$;
 \item[(c)] the maximum convection zone mass extension (hereafter MCZME) during core He-burning s-process;
 \item[(d)] the duration of core He-burning s-process.
\end{itemize}

For all the stellar models of different initial mass and metallicity, the s-process efficiency increases when overshooting is inserted in the evolutionary computations compared with ``no-overshooting'' models, as found in \citetalias{pumo2006} and \citetalias{costa2006} for models with $M_{ZAMS}= 25 M_\odot$ and $Z=0.02$. Moreover an essentially monotonic link between the $f$ value and the s-process efficiency is evident when overshooting is inserted in the evolutionary computations, as witnessed by the fact that all s-process indicators gradually grow when increasing the $f$ value for any set of models. Also evident is a clear trend with both initial mass and metallicity when setting the $f$ value, according to which the s-process efficiency increases when we progressively increase both the mass and the metallicity in their ranges, confirming the results found in other works referring to evolutionary computations without extra mixing processes due to convective overshooting \citep[see e.g.][]{prantzos90,rayethashimoto2000,the2000}. There are only few exceptions in these trends, explainable with a ``relatively'' long duration of the core He-burning s-process (see \citetalias{pumo2010} for details).


\section{Further comments}

These results clearly show the level of uncertainty in the modelling of the weak s-process component due to the lack of a self consistent theory of stellar convection. In particular, current uncertainties due to overshooting do actually give rise up to a factor $\sim 6$ uncertainty in the s-process efficiency; thus, prior to giving a final conclusion on the possible contribution of post-He burning phases to the s-process yields from a quantitative point of view, some additional investigation taking into account stellar evolution uncertainties in addition to the nuclear physics ones, should be performed.

Moreover, as better explained in \citetalias{pumo2010}, this additional investigation may shed light on different open questions linked, for example, to the effective existence of the LEPP process and to the model for the p-process taking place in the type II supernovae O-Ne layers, because the relevant s-nuclei are p-process seeds.


\begin{acknowledgements}
M.L.P. acknowledges the support by the Bonino-Pulejo Foundation.
\end{acknowledgements}

\bibliographystyle{aa}

\end{document}